\documentclass[preprint,12pt]{elsarticle}

\usepackage{graphicx}
\usepackage{dcolumn}
\usepackage{bm}
\usepackage{amsmath}
\usepackage{amssymb}
\newcommand{\beq}{\begin{equation}}
\newcommand{\eeq}{\end{equation}}
\newcommand{\bea}{\begin{eqnarray}}
\newcommand{\eea}{\end{eqnarray}}

\begin{document}

\begin{frontmatter}



\title{Temperature dependence of the mean magnon collision time in a spin Seebeck device}


\author{Vittorio Basso, Alessandro Sola, Patrizio Ansalone and Michaela Kuepferling}

\address{Istituto Nazionale di Ricerca Metrologica, Strada delle Cacce 91, 10135, Torino, Italy}

\begin{abstract}
Based on the relaxation time approximation, the mean collision time for magnon scattering $\tau_c(T)$ is computed from the experimental spin Seebeck coefficient of a bulk YIG / Pt bilayer. The scattering results to be composed by two processes: the low temperature one, with a $T^{-1/2}$ dependence, is attributed to the scattering by defects and provides a mean free path around 10 $\mu$m; the high temperature one, depending on $T^{-4}$, is associated to the scattering by other magnons. The results are employed to predict the thickness dependence of the spin Seebeck coefficient for thin films.
\end{abstract}


\end{frontmatter}


\section{Introduction}

A spin Seebeck device is a bilayer composed by a ferromagnetic insulator (i.e. the ferrimagnet YIG) and a metal with a strong spin Hall effect (i.e. Pt). The spin Seebeck effect is obtained by applying a temperature gradient to the YIG that generates a current of magnetic moment. Part of the magnetic moment current from YIG is injected at the interface into the side metallic layer where it is carried by electrons. The use of Pt as a metallic layer permits to detect the magnetic moment current because of the inverse spin Hall effect that laterally deflects polarized electrons and creates an electric voltage in the transverse direction \cite{Uchida-2010}. In the last decade the spin Seebeck effect has attracted attention as an alternative thermoelectric generator and as a source of a magnetic moment current for spintronic devices without involving a charge current \cite{Uchida-2016,Yu-2017c}. However, to optimize the effect, the underlying physics needs to be appropriately understood. At any finite temperature the saturation magnetization of the ferromagnet is decreased with respect to its spontaneous value because of the thermal excitation of spin waves (magnons). It is believed that the temperature gradient across the ferromagnet forces the diffusion of the thermal magnons in the direction of the gradient, therefore generating a current of magnetic moment \cite{Xiao-2010, Zhang-2012, Rezende-2014}. A crucial point is therefore to understand the transport properties of magnons and their scattering \cite{Boona-2014b}. Experiments of the temperature dependence of the spin Seebeck coefficient \cite{Kikkawa-2015, Guo-2016} revealed non obvious features: in bulk YIG single crystals a peak is observed at temperatures around 70 K. The peak is shifted at higher temperatures and partially suppressed by both decreasing the thickness \cite{Kikkawa-2015} or decreasing the grain size in polycrystalline materials \cite{Uchida-2012, Miura-2017}. The interpretation of this effect has stimulated a debate on the possible sources of the scattering of the magnons: impurities, phonons, other magnons and so on \cite{Uchida-2012, Boona-2014b, Guo-2016}. One of the problems is to disentangle the temperature dependence of the magnon scattering processes from the temperature dependence other parameters such as the conductance and the diffusion length of Pt.

\begin{figure}[htb]
\centering
\includegraphics[width=8cm]{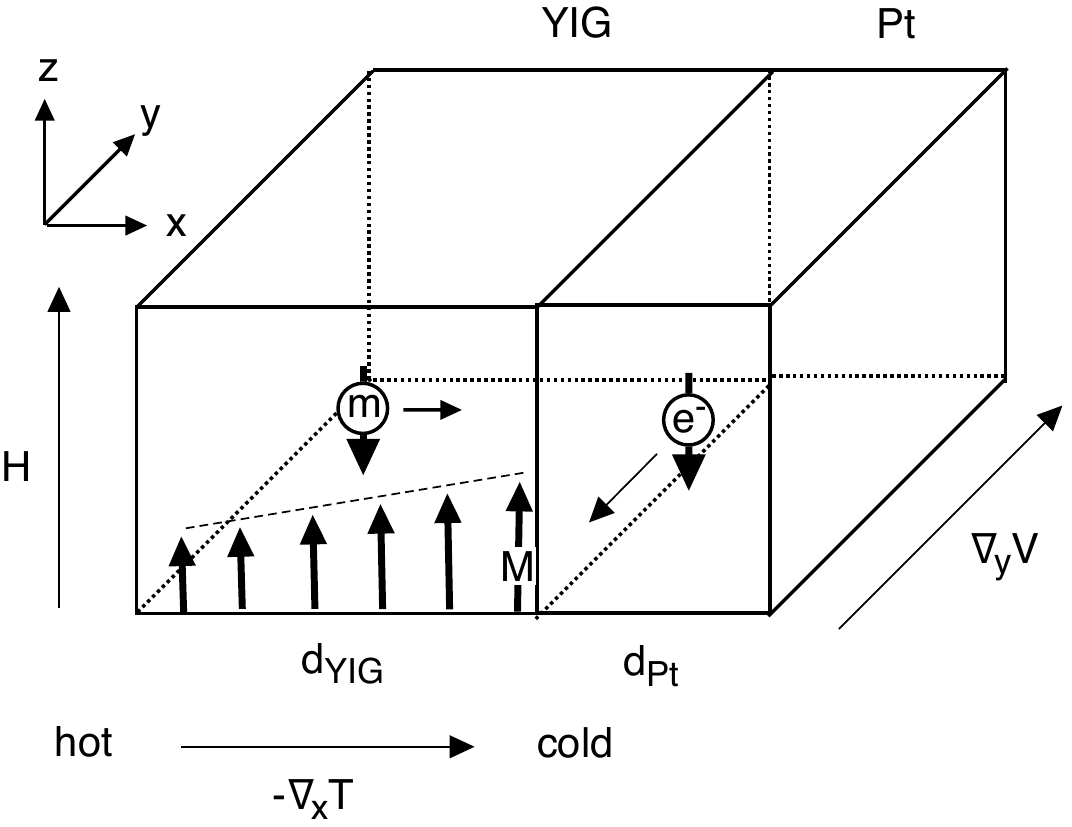}
\caption{Spin Seebeck bilayer composed by a ferromagnetic insulator (i.e. the ferrimagnet YIG) and a metal with a strong spin Hall effect (i.e. Pt). The temperature gradient along $x$ on YIG injects a current of magnetic moment into Pt where it is revealed as a voltage gradient along $y$ because of the inverse spin Hall effect.}
\label{FIG:device}
\end{figure}

Unlike the classical Seebeck effect of metals, in which the ratio voltage over temperature gives the absolute thermo-electric power coefficient, in the spin Seebeck effect the coefficient given by the ratio between the voltage gradient over the temperature gradient (in the geometry of Fig.\ref{FIG:device}) depends not only on the absolute thermomagnetic power coefficient of the YIG, $\epsilon_{\rm YIG}$, and on the spin Hall angle of the platinum, $\theta_{\rm SH}$, but also on the magnetic moment conductances, the diffusion lengths and the thicknesses of the two layers \cite{Basso-2018}. Even if certain parameters can be obtained by independent experiments, the expression of the spin Seebeck coefficient still contains at least two free parameters: $\epsilon_{\rm YIG}$ and the magnetic moment conductivity $\sigma_{\rm M,YIG}$ of the ferromagnet. To simplify the problem in this work we set $\epsilon_{\rm YIG} = -0.956$ T/K as predicted by the Boltzmann approach to the transport of magnons in the relaxation time approximation\cite{Nakata-2015b, Basso-2016b}. Then the magnetic moment conductivity can be easily deduced from the experimental data and the relaxation time $\tau_c(T)$ computed as a function of temperature. $\tau_c(T)$ is an estimate of the typical scattering time and gives insights on the evolution of the scattering processes as a function of the temperature. The main result of this paper is to show that the scattering is composed by two processes. At low temperature the active process changes with temperature as $T^{-1/2}$, a dependence that can be attributed to the scattering by defects. At high temperature it depends strongly on $T$ as $T^{-4}$ a variation that could be associated to the scattering by other magnons.

\section{Spin Seebeck effect}

The spin Seebeck coefficient is given by the ratio between the voltage gradient in platinum and the temperature gradient in YIG (see Fig.\ref{FIG:device})

\beq
S_{\rm SSE} = \frac{\nabla_{y} V_{e}}{\nabla_xT}
\eeq

\noindent To derive a theoretical expression for the spin Seebeck coefficient one has to use a thermodynamic theory describing the transport of the magnetic moment in the ferromagnet caused by the temperature gradient and the transverse electric effect of the inverse spin Hall effect in the metal \cite{Basso-2016}. For both materials one has also to take into account that the magnetic moment is a non conserved quantity and therefore the transport of the magnetic moment is characterized by a diffusion length $l_M$ and by a typical time constant $\tau_M$. The two values and their ratio $v_M = l_M/\tau_M$, a parameter that assumes the meaning of a magnetic moment conductance, are characteristic values of each material (M = Pt, YIG) and can be temperature dependent. From the thermodynamic theory by exploiting the reciprocity of both the intrinsic thermal and magnetic transport in the ferromagnet and the spin Hall effect in the metal \cite{Sola-2019}, one derives the following expression (see appendix A for a derivation)

\beq
S_{\rm SSE} = - \theta_{\rm SH} \left(\frac{\mu_B}{e} \right) \frac{l_{\rm YIG} v_{\rm YIG}}{d_{\rm Pt} v} \epsilon_{\rm YIG}
\label{EQ:SSSE}
\eeq

\noindent where $ \theta_{SH}$ is the spin Hall angle of the metal (the angle for the magnetic moment is opposite with respect to the one for the spin i.e. negative for Pt and positive for W and Ta), $\epsilon_{\rm YIG}$ is the thermomagnetic power coefficient of the ferromagnet, $l_{\rm YIG}$ and $v_{\rm YIG}$ are the diffusion length and the magnetic moment conductance of the ferromagnet. The product $l_{\rm YIG} v_{\rm YIG} = \mu_0\sigma_{\rm M,YIG}$ is proportional to the magnetic moment conductivity of the ferromagnet $\sigma_{\rm M, YIG}$. $v$ is an effective conductance of the bilayer and summarizes the effects of the ratio between the thicknesses of the layers and their own intrinsic diffusion lengths and contains the sum of the effective conductances. Its expression is

\beq
\frac{1}{v} = \frac{f(d_{\rm Pt}/l_{\rm Pt}) f(d_{\rm YIG}/l_{\rm YIG})}{v_{\rm Pt}\tanh(d_{\rm Pt}/l_{\rm Pt})+v_{\rm YIG} \tanh(d_{\rm YIG}/l_{\rm YIG})}
\label{EQ:vp}
\eeq

\noindent where $f(x) = \tanh(x) \tanh(x/2)$. in the case of a thick YIG with $d_{\rm YIG} \gg l_{\rm YIG}$ it is simplified as

\beq
\frac{1}{v} = \frac{f(d_{\rm Pt}/l_{\rm Pt})}{v_{\rm Pt}\tanh(d_{\rm Pt}/l_{\rm Pt})+v_{\rm YIG}}
\eeq

\noindent and if the conductance of the metal is much larger than the one of the ferromagnet, $v_{\rm Pt}\tanh(d_{\rm Pt}/l_{\rm Pt}) \gg v_{\rm YIG}$, it is directly given by the properties of the metal $v = v_{\rm Pt}\coth(d_{\rm Pt}/(2 l_{\rm Pt}))$. 

The aim of this paper is to compute $l_{\rm YIG}$ and $v_{\rm YIG}$ of YIG as a function of temperature by using Eq.(\ref{EQ:SSSE}) and the experimental data for the spin Seebeck coefficient. In Eq.(\ref{EQ:SSSE}) many parameters are known from other experiments. The missing one is the intrinsic parameter $\epsilon_{\rm YIG}$. In this work we employ the result of the transport theory of magnons which predicts a temperature independent constant $\epsilon_{\rm YIG} = -0.956$ T/K \cite{Basso-2016b}. 

\begin{figure*}[htb]
\centering
\includegraphics[width=13cm]{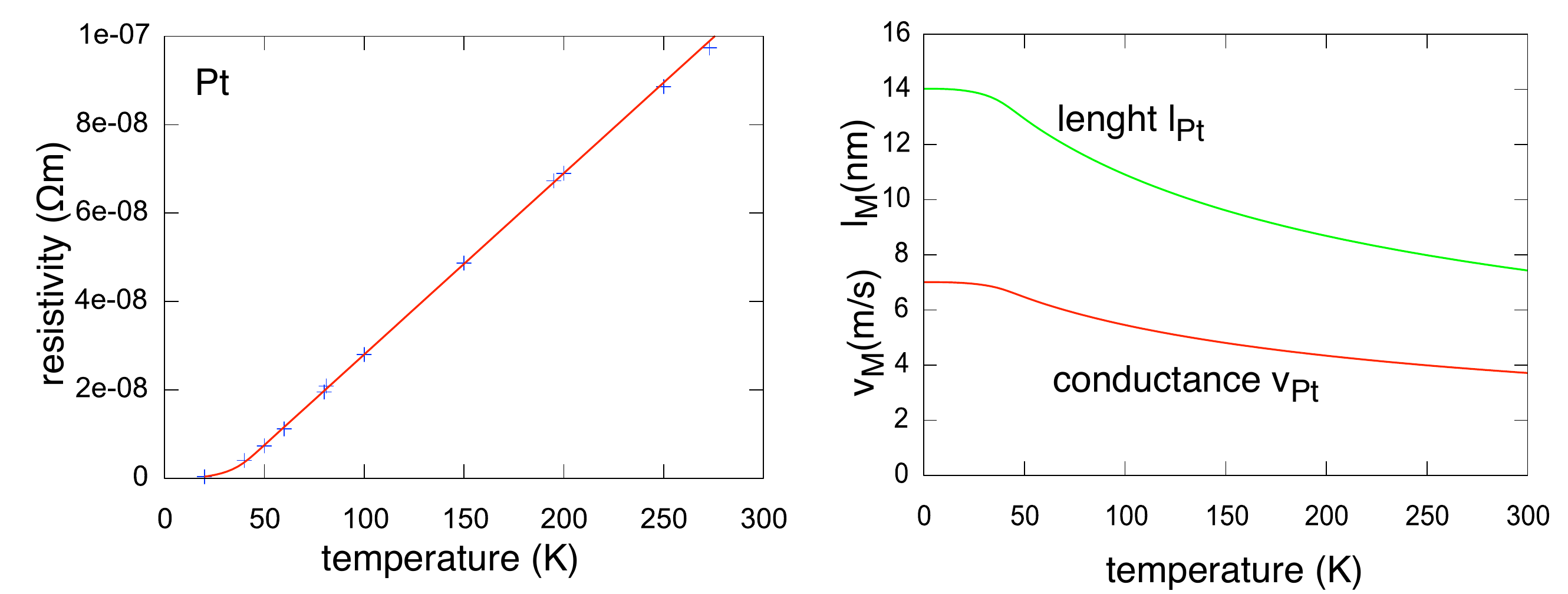}
\caption{Left: standard resistivity of Pt (from Platinum Metals Rev., 28 (4) 164 (1984)) and fitting function $\varrho(T) = x\varrho_{L} +(1-x) \varrho_{H}$ with $\varrho_{L} = 6 \cdot 10^{-14} T^3$, $\varrho_{H} = 4.1 \times 10^{-10} (T-32)$, $x=[1+\tanh[(T-40)/10]]/2$. Right: estimated $l_{\rm Pt} = (\mu_0 \sigma_{\rm M,Pt} \tau_{\rm Pt})^{1/2}$ and $v_{\rm Pt} = l_{\rm Pt}/\tau_{\rm Pt}$ with $\tau_{\rm Pt} = 2.1$ ns, $ \sigma_{\rm M,Pt} = ({\mu_B}/{e})^2 \sigma_{\rm  e,Pt}$,  $\sigma_{\rm e,Pt} = (\varrho(T) + \varrho_{0})^{-1}$ with residual resistivity $\varrho_{0} = 0.43 \times 10^{-7}$ $\Omega$ m. } 
\label{FIG:Pt}
\end{figure*}

We initially simplify the problem by considering a bulk YIG single crystal at constant temperature. For a bulk YIG we have $d_{\rm YIG} \gg l_{\rm YIG}$ and, expecting $v_{\rm Pt} \gg v_{\rm YIG}$, we estimate $v \simeq v_{\rm Pt}\coth(d_{\rm Pt}/(2 l_{\rm Pt}))$. The magnetic moment conductivity of Pt is $ \sigma_{\rm M,Pt} = ({\mu_B}/{e})^2 \sigma_{\rm  e,Pt}$. Then, with $\sigma_{\rm e,Pt} = 6.4\times 10^{6}$ $\Omega^{-1}$m$^{-1}$ we obtain $\mu_0 \sigma_{\rm M,Pt} = 2.6\times 10^{-8}$ m$^{2}$s$^{-1}$. As $\mu_0 \sigma_{\rm M,Pt}=l_{\rm Pt}v_{\rm Pt}$, with $l_{\rm Pt} = 7.3$ nm \cite{Wang-2014}, we get $v_{\rm Pt} = 3.5$ m/s, $\tau_{\rm Pt} = 2.1$ ns and finally $v \simeq 11.2 \,\, \mbox{m/s}$.  The spin Hall angle is taken as $\theta_{\rm SH}=-0.1$ \cite{Wang-2014}. From recent experiments at room temperature with a bulk YIG single crystal ($d_{\rm YIG} = 0.5$ mm)  we found $S_{\rm SSE} = - 4.8 \times 10^{-7}$ V K$^{-1}$ \cite{Sola-2019}. Then we compute the only missing parameter, the product $l_{\rm YIG} v_{\rm YIG}$, resulting $5.3 \times 10^{-9} \, \mbox{m$^2$/s}$. By estimating the time constant of the YIG as $\tau_{\rm YIG} = (\mu_0\gamma_L M_0 \alpha)^{-1}$ where $\alpha$ is the damping, we get $\tau_{\rm YIG} = 1\times 10^{-6}$ s with $\alpha = 2.5 \times 10^{-5}$ ($M_0 =1.95\times 10^{5}$ A/m is the spontaneous magnetization of YIG). Therefore, at room temperature one finds a diffusion length $l_{\rm YIG} \simeq 70 \,\, \mbox{nm}$ and a conductance $v_{\rm YIG} \simeq 0.07 \, \mbox{m/s}$. It must be remarked that the diffusion length computed in this way is very similar to the one deduced from the YIG thickness dependence in Ref.\cite{Kehlberger-2015}. 

\begin{figure*}[htb]
\centering
\includegraphics[width=13cm]{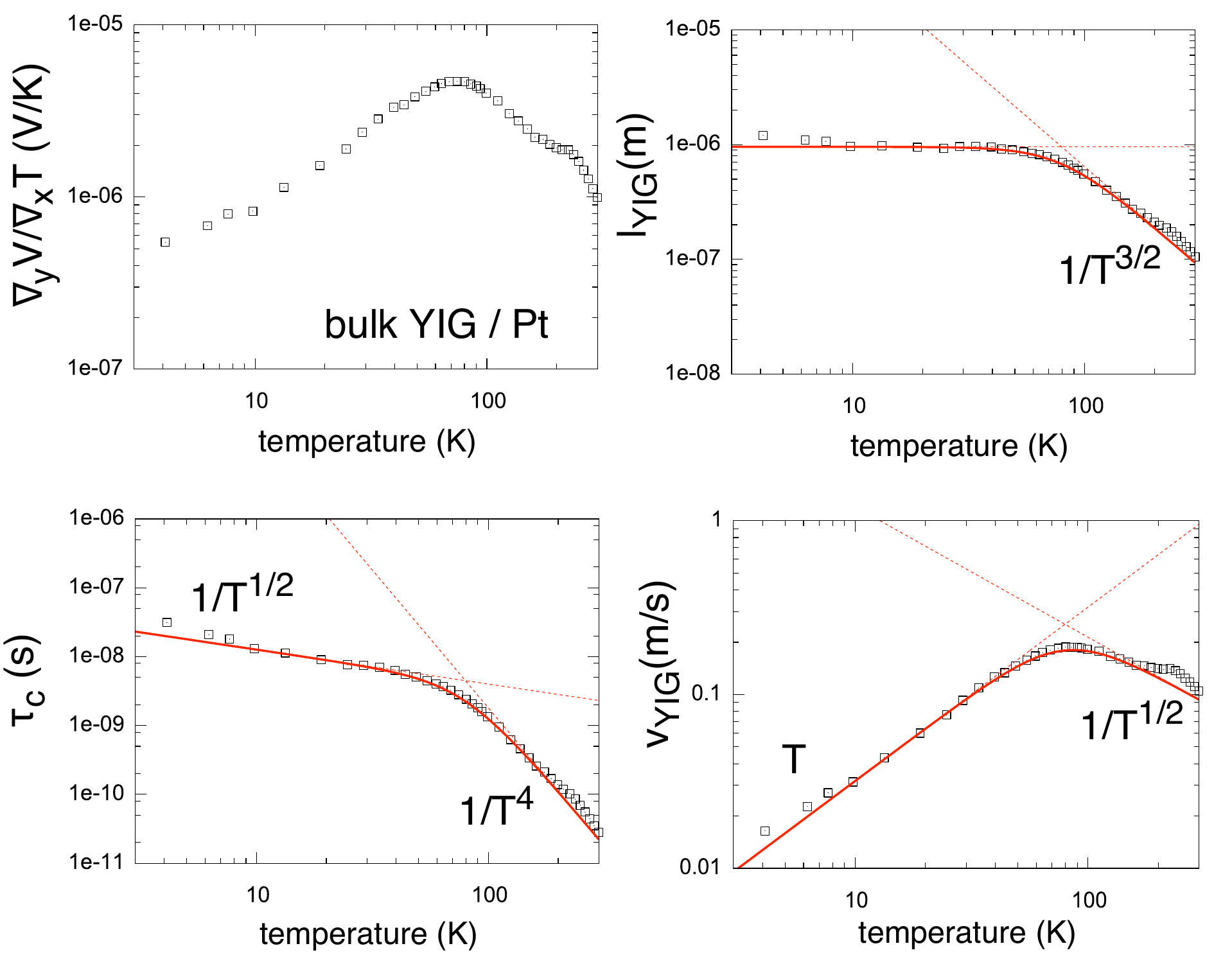}
\caption{Top left: spin Seebeck coefficient for a bulk (1 mm) YIG from Ref.\cite{Kikkawa-2015}. Top right: diffusion length $l_{\rm YIG}(T)$. Bottom right: magnetic moment conductance $v_{\rm YIG}(T)$. Bottom left: relaxation time $\tau_{c}$. Points are computed from the data of Ref.\cite{Kikkawa-2015}. Dashed lines are fitted behavior in terms of the powers of temperature marked on the graphs. Full line $\tau_{c}^{-1} = \tau_{c,L}^{-1}+\tau_{c,H}^{-1}$ with $\tau_{c,L} = 1.8 \times 10^{-9} (T/T_m)^{-1/2}$ s, $\tau_{c,H} = 3.4 \times 10^{-12} (T/T_m)^{-4}$ s. } \label{FIG:SSE}
\end{figure*}

The same calculation can be performed by considering the temperature dependence of the spin Seebeck coefficient for the YIG single crystal ($d_{\rm YIG} = 1$ mm) of Ref.\cite{Kikkawa-2015} (we take from now on the $S_{\rm SSE}$ as an absolute value without the minus sign given by our choice of the reference system). For platinum, the spin Hall angle is not expected to change significantly with $T$ \cite{Isasa-2015}, therefore we assume $\theta_{SH}=-0.1$, while the temperature dependence of the diffusion length and of the conductance of Pt are estimated on the basis of the temperature dependence of the resistivity of Pt (see Fig.\ref{FIG:Pt} left) as shown in Fig.\ref{FIG:Pt} right. For YIG, $\epsilon_{\rm YIG}$ is expected from the diffusion theory to be temperature independent \cite{Basso-2016b}, while the temperature dependence of the time constant $\tau_{\rm YIG}$ is attributed to the temperature dependence of $\alpha$. In Ref.\cite{Maier-Flaig-2017} it was found that $\alpha$ is approximately linear with $T$. With these assumptions one derives $l_{\rm YIG}(T)$ and $v_{\rm YIG}(T)$ shown in Fig.\ref{FIG:SSE} right. Again here it is remarkable that, by decreasing the temperature, the diffusion length reaches a plateau at around $1 \mu$m that matches the low temperature estimate of Ref.\cite{Kehlberger-2015}.

Now, having the temperature dependence of $l_{\rm YIG}(T)$ and $v_{\rm YIG}(T)$ it is possible to employ Eq.(\ref{EQ:SSSE}) with Eq.(\ref{EQ:vp}) to compute the spin Seebeck coefficient as a function of both YIG thickness and temperature. The result is shown in Fig.\ref{FIG:Thickness} left. The set of curves fully captures the phenomenology of the experimental curves of Refs.\cite{Kikkawa-2015} and \cite{Guo-2016} but the peak starts to decrease at a YIG thickness ($\sim 2 \, \mu$m) much smaller than the one seen in experiments ($\sim 20 \, \mu$m). To understand more we have therefore to go into the details of the scattering processes.

\section{Magnon collision time}

\begin{figure*}[htb]
\centering
\includegraphics[width=13cm]{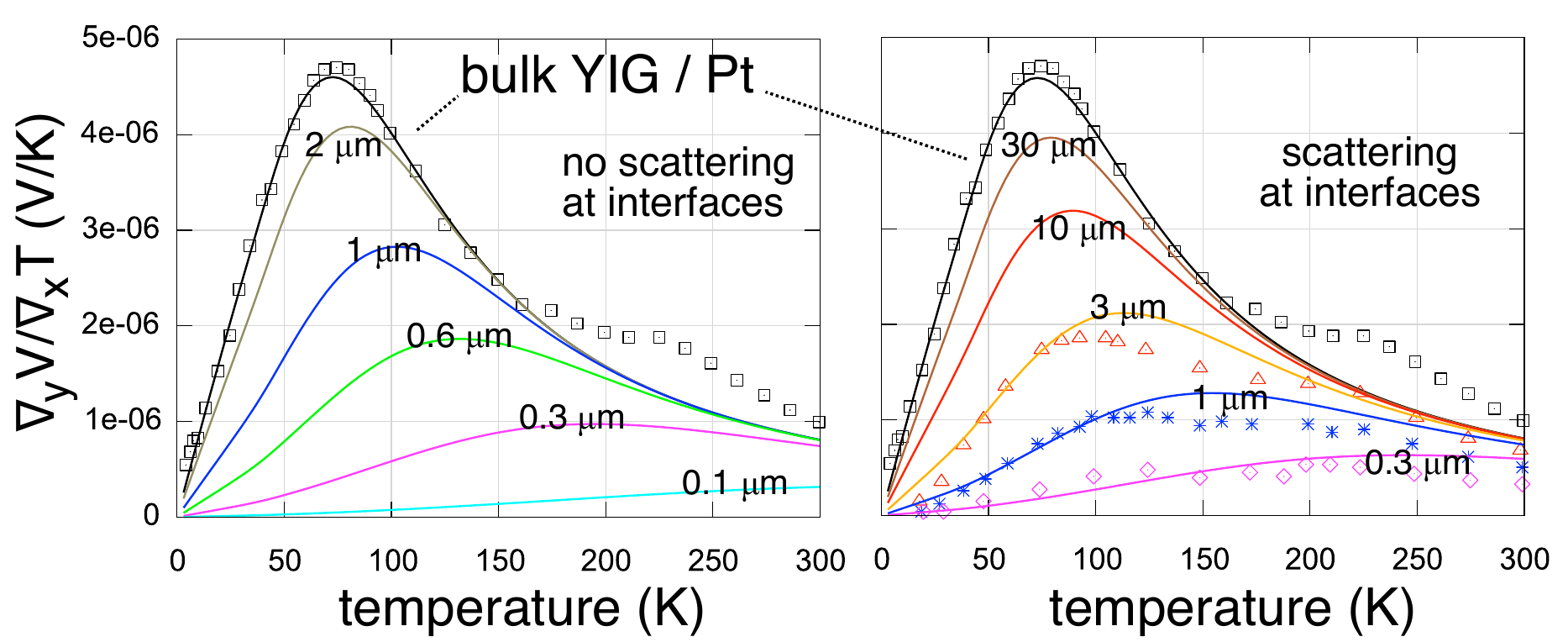}
\caption{Spin Seebeck coefficient as a function of thickness and temperature. Lines are theoretical predictions. Left: computed by using the scattering corresponding to the bulk for all thicknesses. Right: including a an additional scattering contribution due to interfaces a bulk. In both cases the points corresponds to the data of the spin Seebeck coefficient YIG from Ref.\cite{Kikkawa-2015} (black squares: 1mm, red triangles 10 $\mu$m, blue stars 1$\mu$m, purple diamonds 0.3$\mu$m).} \label{FIG:Thickness}
\end{figure*}

The previous results have been obtained by fixing the thermomagnetic power coefficient of the ferromagnet $\epsilon_{\rm YIG}$ to the constant -0.956 T/K which is the result of the Boltzmann transport theory in the relaxation time approximation \cite{Nakata-2015b, Basso-2016b}. With the relaxation time approximation one assumes that the diffusing magnons relax to the equilibrium distribution with a typical time constant $\tau_{c}$. From the theory the magnetic moment conductivity is given by

\beq
\sigma_{\rm M, YIG} = \frac{(2\mu_B)^2 n_m \tau_c}{m_m^*}
\eeq

\noindent where $2\mu_B$ is the magnetic moment carried by the magnon, $m_m^*$ is its effective mass $m_m^* = {\hslash^2}/({2D})$, where $D$ is the spin wave exchange stiffness, $\tau_c$ is the relaxation time and $n_m$ is the number of magnons. The number of magnons is expected to change significantly with temperature and $n_m$ is given by

\beq
n_m = \frac{n}{8\pi^{3/2}}\left(\frac{T}{T_m}\right)^{3/2}
\eeq

\noindent where $n$ is the volume density of localized magnetic moments in the system $n= 1/a^3$, $T_m = {D}/({a^2k_B})$ and $k_B$ is the Boltzmann constant. $a$ is typical distance between localized magnetic moments and is related to the spontaneous magnetization by $M_0 = \mu_B /a^3$. For YIG one has $D = 8.6 \times 10^{-40}$ J m$^2$, $m_m^* = 2.5 \times 10^{-28}$ kg (about 280 times the electron mass) and $T_m=480$ K \cite{Stancil-2009}. As the temperature dependence of the magnetic moment conductivity $\mu_0 \sigma_{\rm M,YIG} = l_{\rm YIG} v_{\rm YIG}$ has been estimated from the spin Seebeck coefficient, we can deduce the temperature dependence of the relaxation time $\tau_c(T)$ describing the collision processes. The result is shown in Fig.\ref{FIG:SSE} bottom left. It appears that the scattering of the magnons is the superposition of two mechanisms. If we apply the Matthiessen's rule 

\beq
\frac{1}{\tau_c} = \frac{1}{\tau_{c,L}}+\frac{1}{\tau_{c,H}} 
\eeq

\noindent we can separate two different time constants. At low temperature the active process $\tau_{c,L}$ appears to change with temperature as $T^{-1/2}$ while at high temperature $\tau_{c,H}$ depends strongly on $T$ as $T^{-4}$. The low temperature process can be attributed to the scattering by defects. If we take the velocity of the magnons to vary with temperature as $v_m = \sqrt{k_B T / m^*}$ and compute a mean free path due to intrinsic defects as $\lambda = v_m \cdot \tau_{c,L}$ we get a temperature independent value of $\lambda \simeq$ 10 $\mu$m. We recall that the mean free path was obtained from the bulk data. In the case of thin films with $d_{\rm YIG} < \lambda$ therefore we have to introduce an additional extrinsic time constant $\tau_{c,L,e}$ which is thickness dependent as $\tau_{c,L,e} = d_{\rm YIG}/v_m$ in order to add the scattering at the interface. The curves calculated with this additional scattering mechanism can be seen in Fig.\ref{FIG:Thickness} right. It must be observed that the plot at the right shows the decrease of the signal at thicknesses much larger with respect to the case of the plot of the left and in better agreement with the experiments. Considering that all the parameters have been derived only from the data of the bulk, the agreement is reasonably good. The high temperature time constant $\tau_{c,H}$ decreasing as $T^{-4}$ is exactly what is expected for the magnon-magnon scattering process \cite{Erdos-1965}.

\section{Conclusions}

In the paper we have performed a study of the temperature dependence of the spin Seebeck coefficient of the YIG/Pt bilayer. We have interpreted the literature data \cite{Kikkawa-2015, Guo-2016} by using the thermodynamic theory of Refs.\cite{Basso-2016, Basso-2016b} in which the magnon transport occurs by diffusion in presence of a gradient of the thermodynamic temperature and in the gradient of the potential for the magnon diffusion. The resulting temperature dependence of the relaxation time describes two sources of scattering: the scattering by defects or interfaces, changing with temperature as $T^{-1/2}$, and mainly active at low temperature and a scattering by other magnons, depending on $T^{-4}$ and active at high temperature.

The classical literature investigating the contributions of magnons to the thermal conductivity of ferromagnetic insulators, has always considered that the magnon-magnon process was very difficult to identify. This is because it is relevant only at high temperatures, a range in which the contribution of magnons to the transport of heat is irrelevant with respect to the phonon one. The spin Seebeck effect represents an case for the detailed study of magnon-magnon scattering because it reveals the transport of magnetic moment. In reference to the recent literature claiming a magnon-phonon drag effect \cite{Adachi-2010}, the results obtained in this study are indicating that the magnon-phonon process as a main source of scattering can be excluded. However it cannot be a priori excluded that, especially in materials with a strong magnetoelastic interaction, the magnon-phonon processes, by means of a phonon drag effect, could give an enhancement of the effective $\epsilon_{\rm YIG}$ \cite{Adachi-2010}. From our study it appears that such a mechanism is not necessary to explain the peak of the spin Seebeck signal of YIG. Future works could clarify the limits and possibilities of these two approaches. 

A possible interesting extension of the present approach to the magnon transport is at high temperatures close to the Curie point. This extension is not straightforward because the spin wave spectrum (especially the long wavelengths) develops over a background saturation magnetization $M_s$ which is less than the spontaneous $M_0$ and such a decrease is due to the spin wave themselves (especially the short wavelengths). Possible advancements of a Boltzmann transport approach are possible in terms of a self consistent approach as suggested by Robert Brout with his random phase approximation technique \cite{Brout-1965}. 

\appendix
\section{Spin Seebeck coefficient}

The spin Seebeck coefficient of Eq.(\ref{EQ:SSSE}) is calculated as follows \cite{Basso-2016}. The transverse voltage gradient in Pt is given by the formula

\beq
\nabla_{y} V_{e} = \frac{1}{\sigma_{e,Pt}} \theta_{SH} \left(\frac{e}{\mu_B} \right) < j_{M}>_{Pt}
\eeq

\noindent where $< j_{M}>_{Pt}$ is the average magnetic moment current in Pt. By solving the diffusion equation for the magnetic moment in Pt, $< j_{M}>_{Pt}$ is estimated as a function of the magnetic moment current injected at the interface by the ferromagnet, $j_{M,0}$, by assuming that the current at the other end of Pt is zero. The result is

\beq
< j_{M}>_{Pt} = \frac{l_{Pt}}{d_{Pt}} \tanh(d_{Pt}/(2 l_{Pt}))j_{M,0}
\eeq

\noindent Similarly by solving the diffusion equation also in the YIG and imposing the boundary conditions between the metal and the ferromagnet, $j_{M,0}$ is computed as a function of the magnetic moment current source $j_{MS}$ due to the intrinsic spin Seebeck effect in YIG. $j_{M,0}$ is given by the current divider equation

\beq
j_{M,0} = \frac{v_{Pt}\tanh(d_{Pt}/l_{Pt})}{v_{Pt}\tanh(d_{Pt}/l_{Pt})+v_{M} \tanh(d_{M}/l_{M})}j_{MS}
\eeq

\noindent where the expressions of the type $v_M \tanh(d_{M}/l_{M})$ are the effective conductances (depending on the ratio $d_M/l_M$) and $v_M$ is the intrinsic magnetic moment conductance which is a property of each material. For a ferromagnet of finite thickness one has that the magnetic moment source is

\beq
j_{MS}= - \tanh(d_{\rm YIG}/(2l_{\rm YIG}))\tanh(d_{\rm YIG}/l_{\rm YIG})\epsilon_{\rm YIG}\sigma_{\rm M,YIG} \nabla_xT
\eeq

\noindent By joining the previous equations one obtains expression (\ref{EQ:SSSE}) with $v$ given by Eq.(\ref{EQ:vp}).

\end{document}